# Towards Design Methodology of Efficient Fast Algorithms for Accelerating Generative Adversarial Networks on FPGAs

Jung-Woo Chang*, Saehyun Ahn*, Keon-Woo Kang, and Suk-Ju Kang
Department of Electronic Engineering, Sogang University, Seoul, South Korea
{zwzang91, hh585, kkw0526, sjkang}@sogang.ac.kr

**Abstract** – Generative adversarial networks (GANs) have shown excellent performance in image and speech applications. GANs create impressive data primarily through a new type of operator called deconvolution (DeConv) or transposed convolution (Conv). To implement the DeConv layer in hardware, the state-of-the-art accelerator reduces the high computational complexity via the DeConv-to-Conv conversion and achieves the same results. However, there is a problem that the number of filters increases due to this conversion. Recently, Winograd minimal filtering has been recognized as an effective solution to improve the arithmetic complexity and resource efficiency of the Conv layer.

In this paper, we propose an efficient Winograd DeConv accelerator that combines these two orthogonal approaches on FPGAs. Firstly, we introduce a new class of fast algorithm for DeConv layers using Winograd minimal filtering. Since there are regular sparse patterns in Winograd filters, we further amortize the computational complexity by skipping zero weights. Secondly, we propose a new dataflow to prevent resource underutilization by reorganizing the filter layout in the Winograd domain. Finally, we propose an efficient architecture for implementing Winograd DeConv by designing the line buffer and exploring the design space. Experimental results on various GANs show that our accelerator achieves up to 1.78×~8.38× speedup over the state-of-the-art DeConv accelerators.

## I. INTRODUCTION

Recently, deep convolutional (Conv) neural networks have demonstrated breakthroughs in a wide range of areas including object detection, classification, and speech recognition [1, 2]. However, since these applications heavily rely on labeled training data, they require a lot of human effort to generate the labeled data. To solve this problem, generative adversarial networks (GANs), which generate new samples from high-dimensional data distributions, are recognized as an attractive solution [3]. Usually, GANs consist of generators and discriminators that compete to learn a high dimensional data distribution. After training each model, generators, which are mainly composed of deconvolutional (DeConv) layers, also called transposed Conv layers, produce synthetic data similar to the original training data [4, 5, 6, 7]. Recently, various hardware accelerators have been proposed to improve the computational complexity of Conv and DeConv operations. Among them, FPGA-based designs are recognized as a promising solution because of their high performance, energy efficiency, and fast re-configurability [8].

The Winograd algorithm can greatly improve resource efficiency by reducing the number of multipliers. However, DeConv, which performs a different type of mathematical operation, has not been applied to fast algorithms. Let $H_I$ and $W_I$ denote the height and width of the input feature map, and $H_O$ and $W_O$ denote the height and width of the output feature map. There are three types of approaches to implement the DeConv layer with hardware accelerators. First, the standard DeConv creates two $K_D \times K_D$ output blocks from input pixels, as shown in Fig. 1(a). However, an overlapping sum problem occurs where output blocks generated from neighboring input pixels overlap each other [9]. This problem interferes with the dataflow of the DeConv accelerator. Second, there are zero padded DeConv-based accelerators [10, 11, 12] to avoid the overlapping sum problem as shown in Fig. 1(b). However, inserting zero values causes very inefficient implementation due to the larger loop dimension [13]. Third, our previous works [14, 15, 16] proposed a method of transforming the DeConv layer into the Conv layer (TDC), which is a spatial transform of the DeConv layer, as shown in Fig. 1(c). This method improves throughput by increasing the parallelism of DeConv and produces the same result as the standard DeConv.

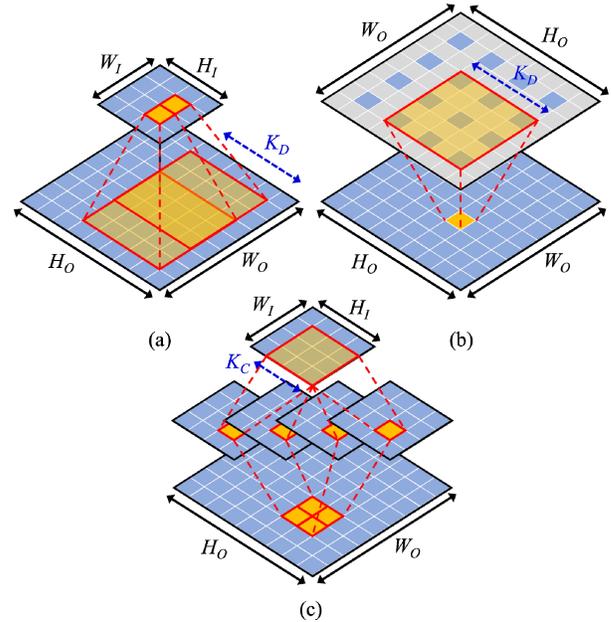

Fig. 1. Three types of DeConv operations: (a) standard DeConv [9] (b) zero padded DeConv [10, 11, 12] (c) TDC-based DeConv [14, 15, 16].

In this paper, we identify that the TDC-based DeConv can be combined with the Winograd algorithm and significantly improve its performance. This is because the TDC-based DeConv can perform the same operation as Conv operation and the original $K_D \times K_D$ kernel is converted to the 3×3 or 2×2 kernel. To maximize the benefits from both conversion methods, we propose an efficient Winograd GAN architecture that leverages the well-optimized designs.

The main contributions of this paper are listed as follows:
● A new class of fast algorithm is proposed for DeConv layers using Winograd minimal filtering. The algorithm computes low complex DeConv using small tiles.

---
* indicates equal contribution

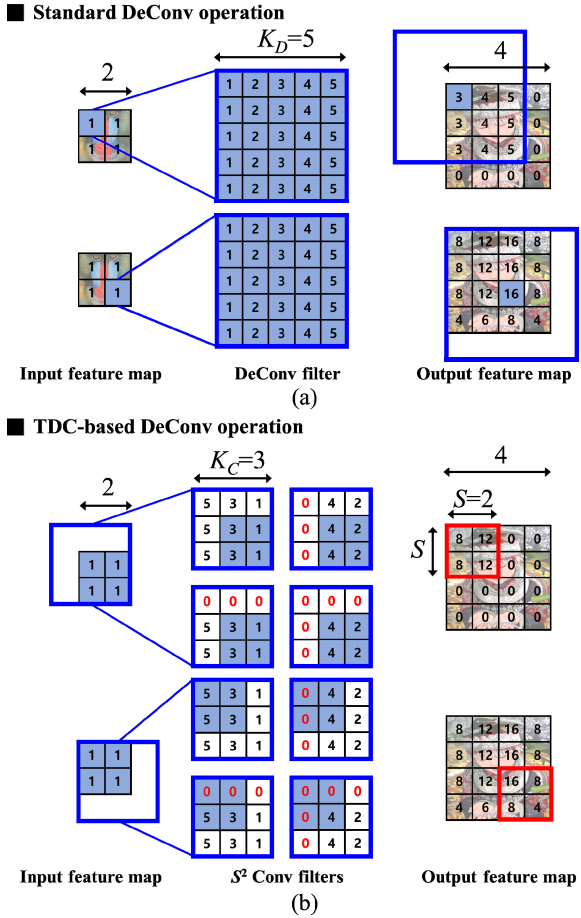

Fig. 2. Examples of DeConv implementations: (a) standard DeConv [9] and (b) TDC-based DeConv [14, 15, 16]. The red bounding box indicates the $S \times S$ output block.

- A novel dataflow is proposed to prevent resource underutilization by reorganizing the filter layout in the Winograd domain. Through this dataflow, the vector-level sparsity, which exists in the transformed filters, is utilized.
- An efficient architecture is presented by designing the novel line buffer and exploring the possible design space for the efficient implementation.

## II. BACKGROUND

### A. Deconvolution Operation

In GANs, DeConv layers are used to increase the amount of input data. As shown in Fig. 2(a), each pixel in the input feature map is expanded with the trained DeConv filter to generate an output block with $K_D \times K_D$ pixels. The output block is then accumulated in the output feature map. However, the final output is not determined until all neighboring blocks are created. Therefore, in the accelerator implementation, there is a serious dataflow problem that requires the output tiles already stored in the off-chip memory to be reloaded.

The TDC method can solve this dataflow problem. As depicted in Fig. 2(b), the input block creates output pixels by computing Conv with the trained $K_C \times K_C$ filter. There are $S^2$ numbers of filters, and each filter creates different output pixels, where $S$ is the stride of the DeConv layer. Using the TDC method, there is no dependency between the inputs when generating each output in the $S \times S$ output blocks. Fig. 2 shows that the results of standard DeConv and TDC-based DeConv are identical. Therefore, the TDC method can improve the data reuse efficiency and the data throughput by creating $S \times S$ output blocks from the same input block. However, there are many zeros in the $S^2$ Conv filters, as shown in Fig. 2(b). Therefore, the consideration for zeros can further improve the efficiency of data processing.

### B. Winograd Algorithm

The Winograd algorithm computes $m$ outputs of an $r$-tap filter via $m+r-1$ multiplications. For clarity, we denote this computation as $F(m, r)$. For example, the computation process for $F(2, 3)$ is derived as

$$Z = [z_0 \ z_1 \ z_2 \ z_3]^T \ f = [x_0 \ x_1 \ x_2]^T \ Y = [y_0 \ y_1]^T,$$

$$\begin{bmatrix} z_0 & z_1 & z_2 \\ z_1 & z_2 & z_3 \end{bmatrix} \begin{bmatrix} x_0 \\ x_1 \\ x_2 \end{bmatrix} = \begin{bmatrix} m_1 + m_2 + m_3 \\ m_2 - m_3 + m_4 \end{bmatrix} = \begin{bmatrix} y_0 \\ y_1 \end{bmatrix}, \quad (1)$$

where

$$m_1 = (z_0 - z_2)x_0 \quad m_2 = (z_1 + z_2)\frac{x_0 + x_1 + x_2}{2}$$
$$m_4 = (z_1 - z_3)x_2 \quad m_3 = (z_2 - z_1)\frac{x_0 - x_1 + x_2}{2}. \quad (2)$$

$Z$, $f$, and $Y$ denote an input, a filter, and an output, respectively.

In this manner, the 1D-Winograd Conv is formulated using three transformation matrices, $A$, $B$, and $G$, as follows,

$$Y = A^T[(Gf) \odot (B^T Z)],$$

$$B^T = \begin{bmatrix} 1 & 0 & -1 & 0 \\ 0 & 1 & 1 & 0 \\ 0 & -1 & 1 & 0 \\ 0 & 1 & 0 & -1 \end{bmatrix} \ G = \begin{bmatrix} 1 & 0 & 0 \\ \frac{1}{2} & \frac{1}{2} & \frac{1}{2} \\ \frac{1}{2} & -\frac{1}{2} & \frac{1}{2} \\ 0 & 0 & 1 \end{bmatrix} \ A^T = \begin{bmatrix} 1 & 1 & 1 & 0 \\ 0 & 1 & -1 & -1 \end{bmatrix}, \quad (3)$$

where $\odot$ is element-wise multiplication.

The number of multiplications is reduced from $m \times r$ to $n$ ($n=m+r-1$) using this algorithm. By nesting the 1D-Winograd algorithm, the computation of 2D-Winograd algorithm is easily described as $F(m \times m, r \times r)$, where the sizes of input tile, output tile, and filter are $n \times n$, $m \times m$, and $r \times r$, respectively. Based on the above representation, 2D-Winograd algorithm is given by:

$$Y = A^T[(GfG^T) \odot (B^T ZB)]A. \quad (4)$$

In the Winograd domain, the number of multiplications required to create the output tile with $m \times m$ pixels is $n^2$. In contrast, the spatial Conv requires $m^2 \times r^2$ multiplications. As a result, the Winograd algorithm greatly leads to a reduction in the number of multiplications. In this paper, the uniform Winograd size of $F(2 \times 2, 3 \times 3)$ is used for all DeConv layers.

## III. PROPOSED APPROACH

### A. Winograd Deconvolution

Table I shows a description of the various GAN models. As shown in Table I, the kernel sizes for various DeConv layers in generative model, $K_D \times K_D$, have 5×5, 4×4, and 3×3 pixels. By applying the TDC method, we convert the $M \times N$ DeConv layer to the $S^2 \times M \times N$ Conv layer, where $M$ and $N$ are the number of output and input feature maps, respectively. Table I shows the widths of the new kernel size, $K_C$. Since $K_C$ is 2 or 3, the newly created Conv layer can be applied to the Winograd minimal filtering algorithm.

Fig. 3 shows the overall process of Winograd DeConv according to the kernel size of DeConv layer, $K_D$. For clarity, the weights in the filters and intermediate results have output indices of 0, 1, 2, and 3 indicating the location of the output pixels. First, $S^2$ kinds of Conv filters are moved to the Winograd domain through a transform matrix $G$. After the computation, zero-valued weights are remained at fixed positions in the transformed filters with a different sparsity pattern. When these new types of transformed filters are implemented in FPGAs, existing Winograd accelerators [17, 18, 19] operate on all weights in $n \times n$ transformed filters, which results in serious resource underutilization. In addition, since the load varies depending on the type and size of the filters, the computational process must adaptively change. Thus, a new dataflow is required to solve this problem.

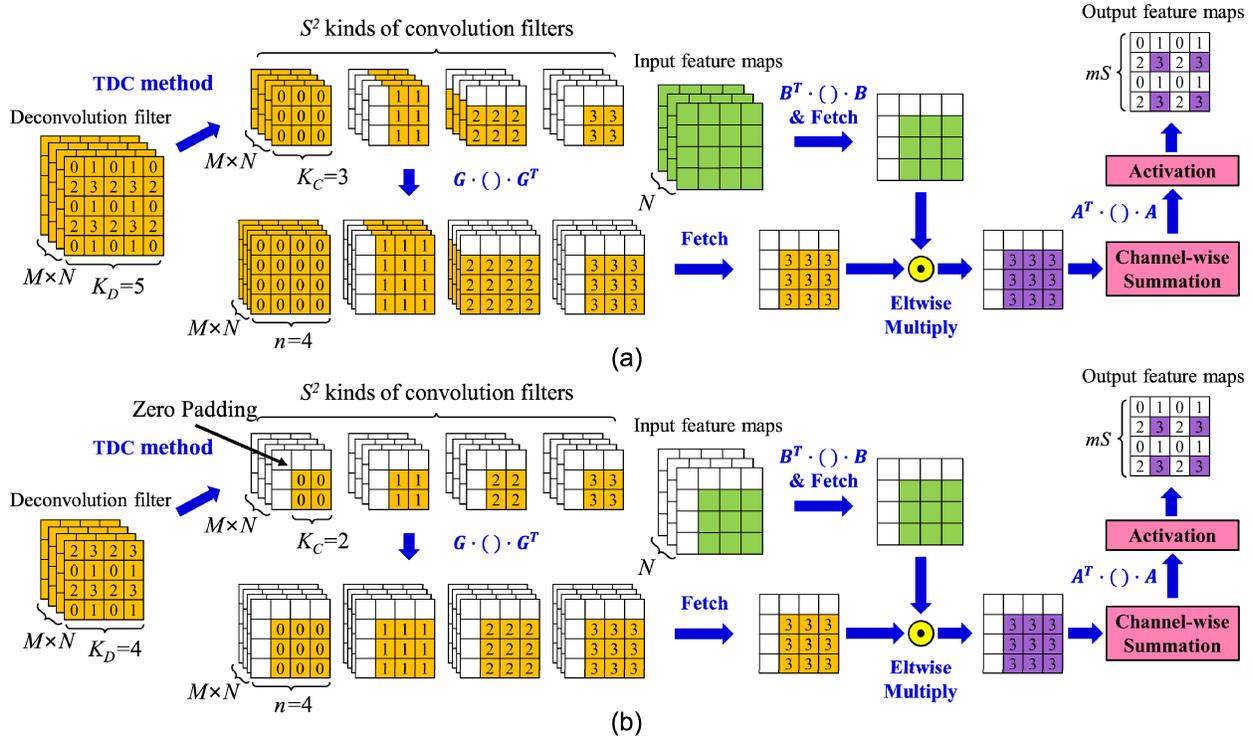

Fig. 3. Combination of TDC method and Winograd algorithm for resource efficiency when kernel size of the DeConv layer, $K_D \times K_D$, is (a) 5×5 and (b) 4×4, respectively. In this figure, $S$ is set to 2, which is typical value in GANs.

The input tiles are moved to the Winograd domain using transform matrix $B$. Transformed filters and input tiles should operate on element-wise multiplication except for the position of zero-valued weights. Then, each of intermediate results accumulates with previous results until all feature maps are processed. After channel-wise summation, we process the inverse transform with matrix $A$. In this process, the latency of the inverse transform can be further reduced by skipping operations in proportion to the number of zero-valued outputs compared to conventional accelerators [17, 18, 19]. Finally, $mS \times mS$ output blocks are created. This is because each filter creates an $S \times S$ output block and simultaneously generates an $m \times m$ output tile via the Winograd algorithm. Fig. 4 shows the number of reduced multiplications required for each DeConv method for various GAN models. Among these methods, zero padded DeConv requires the largest number of computations because it convolves on the up-scaled feature maps with the large kernel size. In contrast, Winograd DeConv requires the smallest number of multiplications because it takes advantage of the combination of the two orthogonal methods, the TDC

TABLE I
VARIOUS GAN MODELS DESCRIPTION

| Name | Year | Generative Networks | | DeConv | | |
| --- | --- | --- | --- | --- | --- | --- |
| | | #_Conv | #_DeConv | $K_D$ | $S$ | $K_C$ |
| DCGAN [4] | 2015 | – | 4 | 5 | 2 | 3 |
| ArtGAN [5] | 2017 | – | 4 | 4 | 2 | 2 |
| | | – | 1 | 3 | 1 | 3 |
| DiscoGAN [6] | 2017 | 5 | 4 | 4 | 2 | 2 |
| GP-GAN [7] | 2019 | – | 4 | 4 | 2 | 2 |

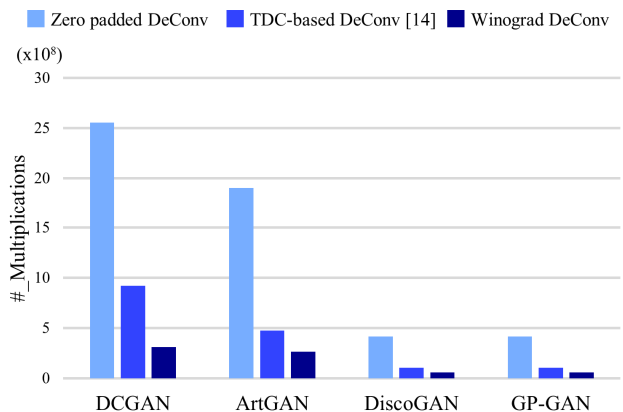

Fig. 4. Total number of reduced multiplications in DeConv layers of various GAN models.

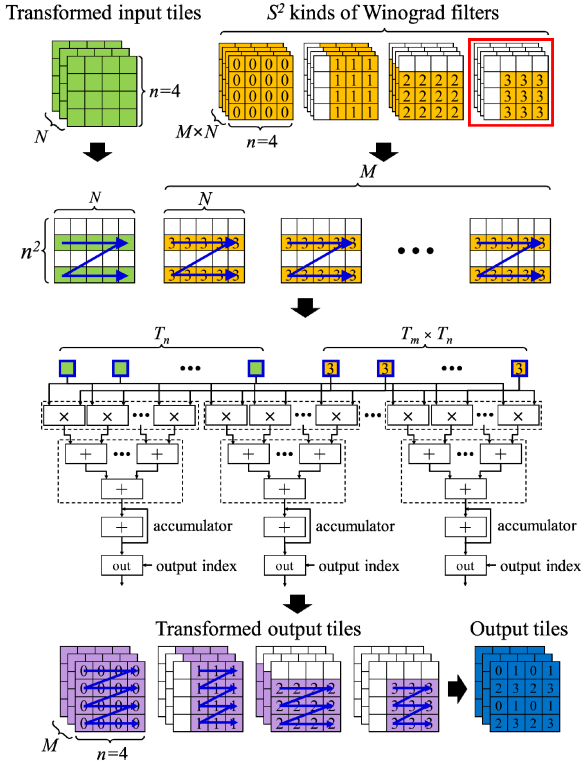

Fig. 5. Overview of Winograd DeConv dataflow.

method and the Winograd minimal filtering algorithm.

### B. Dataflow Optimization

Fig. 5 shows the dataflow of Winograd DeConv. We first fetch the transformed input tiles from the memory. However, each transformed filter has a different ratio of zeros. To solve this problem, we rearrange the transformed input tiles into a matrix of size $n^2 \times N$. Similarly, we reorganize different types of Winograd filter into $M$ matrices of size $n^2 \times N$. In the reordered filters, there is a vector-level sparsity with the same indices of rows due to a regular sparsity pattern. Fig. 6 shows the computation between the reordered tiles and filters depending on the type of filters. First, there is a filter type that does not have any sparsity. In this case, the accelerator does not have the performance benefit from the rearrangement. On the other hand, the second type of filter has vector level sparsity in $n$ rows. Moreover, the third type of filter has the largest number of zeros by having vector-level sparsity in $2n-1$ rows. In these two cases, we have a significant performance enhancement by reducing the idle cycles of an accelerator. In particular, when $K_D$ is 4, as shown in Fig. 3(b), all transformed filters can operate in the Case 3.

As shown in Fig. 5, the accelerating engine outputs $T_m \times T_n$ intermediate results. $T_m$ and $T_n$ are the tile factors of the output and input feature maps. Our accelerator can skip the zero-operand multiplications, so it only outputs non-zero results. In this manner, we eliminate the computations on zero-valued outputs in the inverse transform process.

## IV. ARCHITECTURE DESIGN

### A. Architecture Overview

Fig. 7 shows an overview of the Winograd DeConv architecture, consisting of several processing elements (PEs) and input and output line buffers. First, pre-PE handles the window selection process via the value of $K_C$. Then, pre-PE fetches and transforms the input tiles, and rearranges them with transformed filters. There is an energy overhead between the on-chip buffers in rearranging the transformed input tiles into a matrix of size $n^2 \times N$. Moreover, to find the inputs to compute the multiplication with non-zero weights in the $n^2 \times N$ matrix, additional logic elements are required to determine the inputs according to the values of the output indexes. Then, the reordered inputs and filters are moved to com-PEs in the accelerating engine and make transformed output tiles as shown in Fig. 7. Finally, post-PE converts these tiles into the spatial domain using the sparse inverse transform.

### B. Line Buffer Design

In our design, $S^2$ kinds of Conv filters create $mS \times mS$ output block by computing with $n \times n$ input tiles. Thus, $(n-m) \times n \times S^2$ data reuse can be obtained between two neighboring tiles. Since we cannot store all feature maps and intermediate data in an on-chip buffer, we employ the line buffer that is designed with a simple dual-port mode [20] to overlap the data transfer time between PEs and the computation time between inputs and filters. Next, we should consider the memory to store the incoming inputs while the windows slide the input and output buffers. As shown in Fig. 7, we store $(n+m)$ lines of the $T_n$ input feature maps in the input buffer and $2 \times mS$ lines of the $T_m$ output feature maps in the output buffer, respectively.

### C. Design Space Exploration

We should model the bandwidth between an on-chip buffer and an off-chip memory to determine optimal tiling factors. For efficient ping-pong operation, the data transfer time should be smaller than or equal to the computation time. First, the time for processing $n$ rows in the input buffer, $T_C$, is obtained as follows:

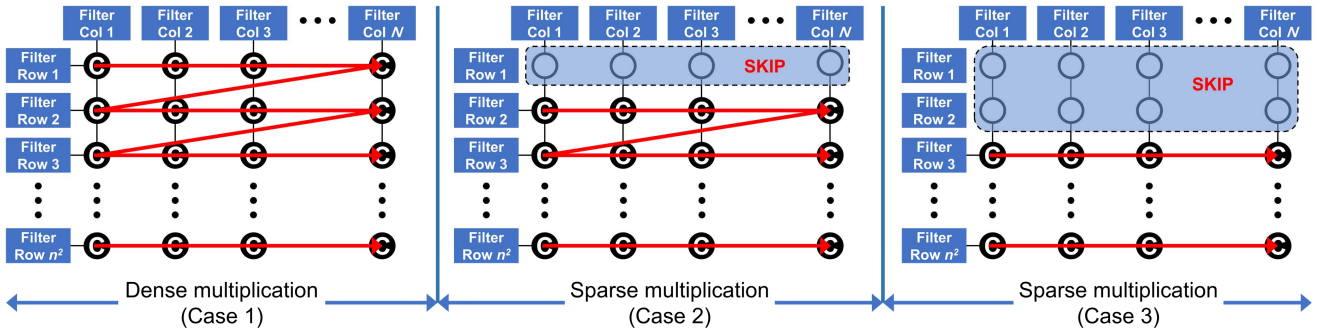

Fig. 6. Computation process according to the vector-level sparsity of each case.

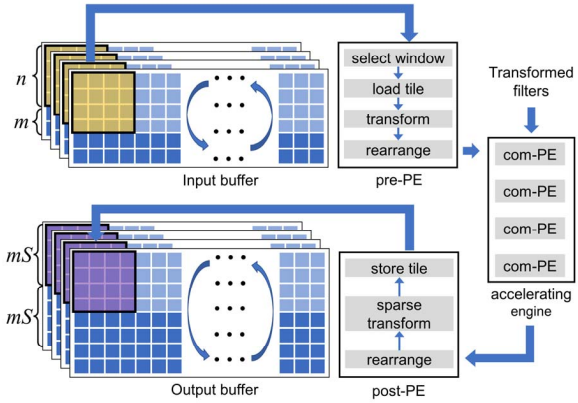

Fig. 7. Block diagram of Winograd DeConv accelerator.

$$T_C = \left\lceil \frac{S^2 M}{T_m} \right\rceil \times \left\lceil \frac{N}{T_n} \right\rceil \times \left\lceil \frac{W_I}{m} \right\rceil \times \frac{C(K_C)}{m^2} \times \frac{1}{freq}, \quad (5)$$

where $C(K_C)$ is 36 (or 49) for $K_C$=2 (or 3).

For most layers, DeConv produces a large amount of output data than input data. Therefore, the data transfer time, $T_D$, is determined based on the output data as follows:

$$T_D = \frac{mS \times W_I \times S^2 M \times n^2}{Bandwidth}. \quad (6)$$

Since $T_D$ should be smaller than $T_C$, the bandwidth requirement can be formulated as follows:

$$Bandwidth = \frac{m^2}{C(K_C)} \times \left\lceil \frac{T_m \times T_n}{N} \right\rceil \times mS \times n^2 \times freq. \quad (7)$$

We denote $T_I$ as the time to fetch the first $n$ rows of input feature maps and filters into on-chip buffer as follows:

$$T_I = \frac{S^2 M \times N \times r^2 + n \times W_I \times N}{Bandwidth / n^2}. \quad (8)$$

As a result, the computational roof can be obtained by the ratio of the total number of operations to the processing time as follows:

$$Computational\ roof = \frac{2 \times S^2 M \times N \times H_I \times W_I \times r^2}{\left\lceil \frac{H_I}{m} \right\rceil \times T_C + T_I}. \quad (9)$$

Enumerating all possible loop orders and tile sizes creates a set of computational roof and bandwidth pairs. We can decide the optimal tiling factors using the cross-layer optimization [21, 22]. We set $T_m$ and $T_n$ to 4 and 128, respectively.

## V. EXPERIMENTAL EVALUATION

### A. Experimental Environment

We conducted the simulations on the Xilinx Virtex7 485T FPGA using single-precision floating-point. The system ran at 100MHz. We used Vivado HLS (v2016.4) to convert the C/C++ code to HDL. Then, we performed the pre-synthesis experiment with C/RTL co-simulation. We also employed the pre-synthesis resource reports for the design space exploration and performance evaluation of multiple constraints. The exported RTL was synthesized and implemented. The off-chip memory was 1 GB DDR3 and the bandwidth between on-chip buffers was 4 GB/s. Table II shows the resources utilization of our design and [14] when implementing DCGAN. Compared to [14], our design required additional operations in pre-PE and post-PE. Thus, we implemented those PEs using LUTs and FFs. We used the same tiling parameters as [14], so the DSP usage was the same. In addition, our design used more BRAMs because we should store more transformed weights in the Winograd domain compared to [14].

TABLE II
RESOURCE UTILIZATION FOR DCGAN

|      | BRAM18K | DSP48E | LUT    | FFs    |
|------|---------|--------|--------|--------|
| [14] | 384     | 2560   | 94264  | 107626 |
| Ours | 520     | 2560   | 142711 | 151395 |

### B. Performance Evaluation

Fig. 8 shows the performance of the zero padded DeConv, the TDC-based DeConv, and the Winograd DeConv in widely used GAN models. Since most GANs consist of DeConv layers for the inference step, we focused on DeConv performance. In DCGAN, the zero padded DeConv should execute filtering on the up-scaled input feature maps using 5×5 kernels. Due to this high computational complexity, there was a technique to skip some of the padded zero activations during the zero padded DeConv [10]. While this approach offered better performance than the zero padded DeConv, it was difficult to avoid redundant computation during the inference. On the other hand, the TDC-based DeConv was 2.79× faster than the zero padded DeConv without any overhead. Our accelerator was 8.38× faster than the zero-padded DeConv and 2.85× faster than the TDC-based DeConv. Next, all layers in ArtGAN could be converted to Conv layers with $K_C$ of 2 through the TDC method. Especially, in the layer where $K_D$ and $S$ were equal to 3 and 1, respectively, the TDC method could not reduce the kernel size, but improved data reuse of inputs by removing an overlapping sum problem [14]. In addition, we improved the performance of the Winograd DeConv by 7.5× and 1.78×, respectively, compared to the zero padded DeConv and the TDC-based DeConv. Similarly, our design showed 7.15× and 1.85× better than two methods in DiscoGAN and GP-GAN.

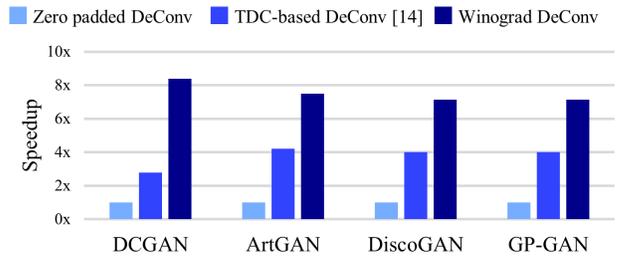

Fig. 8. Performance comparison with other works.

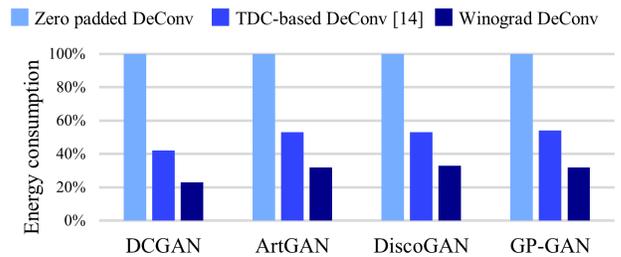

Fig. 9. Energy consumption of DeConv layers compared to zero padded DeConv.

*C. Energy Consumption*

Fig. 9 shows the energy savings achieved by the Winograd DeConv compared to the zero padded DeConv. On average, our design effectively reduced the energy consumption by 3.65× over the zero padded DeConv. The main reason was the difference of the amount of data transfer between the on-chip buffer and the off-chip memory. In addition, the number of the multiplications required was up to 8.16× greater than our design. Compared to the TDC-based DeConv [14], we saw an improvement of 1.74× in terms of energy saving. Although the resource usages of pre-PEs and com-PEs were as large as the tile size, we used a smaller number of multiplications for the fast DeConv. As to the limit of the energy saving, there was a problem of transforming the input tiles that were previously processed in the pre-PE for exploiting the vector-level sparsity.

## VI. RELATED WORK

Despite the cost of having a dimension larger than standard DeConv, many DeConv accelerators have focused on improving the zero-based DeConv to avoid the overlapping sum problem. GANAX [10] presented a unified MIMD-SIMD accelerator by reorganizing the zero patterns to maximize the data reuse and resource utilization. Song et al. [11] proposed a time-multiplexed design to reduce on-chip memory and co-designed an algorithm and architecture to reduce the zero-operand multiplications. To support the operations of Conv and DeConv together, FCN-engine [12] reorganized the conventional Conv accelerator by computing on original input features. ZARA [23] proposed a ReRAM-based accelerator by deforming the zero-operand multiplications and eliminating the huge resource underutilization.

Compared to the zero-based DeConv approaches, the TDC-based DeConv [14] provided a fast and cost-efficient solution for hardware acceleration through DeConv-to-Conv conversion. SDCNN [15] compressed the DCGAN with pruning [24] and designed a sparse GAN accelerator that skips zero-valued weights and employs the TDC-based DeConv. A load balance-aware TDC method was proposed in [16] to increase the efficiency of TDC-based DeConv operations and applied to image super-resolution application for real-time video streaming services.

## VII. CONCLUSIONS

In this paper, we proposed a Winograd DeConv accelerator, which combines the DeConv-to-Conv conversion and the fast algorithm. We first presented a new class of the fast algorithm for DeConv layers of GAN using Winograd minimal filtering. Then, we provided a novel dataflow to prevent resource underutilization by reorganizing the filter layout in the Winograd domain. Finally, we designed an efficient architecture by designing the line buffer and exploring the design space for the efficient implementation. We evaluated our design on Xilinx Virtex7 485T FPGA and results showed that our accelerator achieved 1.78×~8.38× higher throughput over the state-of-the-art accelerators.

## ACKNOWLEDGMENTS

This research was supported by a grant (19PQWO-B153369-01) from Smart road lighting platform development and empirical study on test-bed Program funded by Ministry of the Interior and Safety of Korean government, MSIT (Ministry of Science and ICT), Korea, under the ITRC (Information Technology Research Center) support program (IITP-2019-2018-0-01421) supervised by the IITP (Institute of Information & communications Technology Planning & Evaluation), and Korea Electric Power Corporation. (Grant number: R17XA05-28)